\documentclass{desyproc}

\begin{document}
%------------------------------------
\title{Any Light Particle Search II - Status Overview}

%for single authors the superscripts are optional
\author{{\slshape  No\"emie Bastidon for the ALPS II collaboration,}\\[1ex]
University of Hamburg, Hamburg, Germany}

% if the proceedings are available online (e.g. at Indico)
% please enter the contribution ID or file_name below for the DOI
%\contribID{32}
\contribID{familyname\_firstname}

% TO THE CONFERENCE EDITORS: 
% please update the following information      
% before sending the template to the authors
\confID{11832}  % if the conference is on Indico uncomment this line
\desyproc{DESY-PROC-2015-02}
\acronym{Patras 2015} % if you want the Acronym in the page footer uncomment this line
\doi  % if there is an online version we will register DOIs

\maketitle

\begin{abstract}
The Any Light Particle Search II (ALPS II) experiment (DESY, Hamburg) 
searches for photon oscillations into Weakly Interacting Sub-eV Particles (WISPs). 
This second generation of the ALPS light-shining-through-a-wall (LSW)
experiment approaches the finalization of the preparation phase before ALPS IIa
(search for hidden photons). In the last years, efforts have been put for the
setting up of two optical cavities as well as characterization of a
single-photon Transition-Edge Sensor (TES) detector. In the following, we put
some emphasis on the detector development. In parallel, the setting up of ALPS
IIc (search for axion-like particles), including the unbending of 20 HERA
dipoles, has been pursued. The latest progress in these tasks will be
discussed.  \end{abstract}

\section{Introduction}

The Any Light Particle Search II (ALPS II) experiment (DESY, Hamburg) searches
for photon oscillations into light fundamental bosons (e.g., axion-like
particles, hidden photons and other WISPs) by shining light through a wall \cite{Bahre2013}. The aimed sensitivity increase for the coupling strength of
axion-like particles to photons of the experiment is of 3000 compared to ALPS
I. Such improvements are due to the increase of the magnets length, to two
optical cavities as well as to the replacement of the single-photon detector.
Indeed, the ALPS experiment sensitivity to the conversion of photons into
axion-like particles depends on various parameters and is expressed as

\begin{eqnarray}
S(g_{a\mu})&\propto&(\frac{1}{BL})(\frac{DC}{T})^{\frac{1}{8}}(\frac{1}{\eta\dot{N}_\mathrm{Pr}\beta_\mathrm{PC}\beta_\mathrm{RC}})^{\frac{1}{4}}\nonumber
\end{eqnarray}

with a strong dependency on the magnetic length $L$ and field $B$. The effect
of the optical setup depends on $\dot{N}_\mathrm{Pr}$, the number of injected
photons as well as on $\beta_\mathrm{PC}$ and $\beta_\mathrm{RC}$, the power build-ups of the
production (PC) and regeneration cavities (RC). Finally, the reached sensitivity depends
on the chosen detector's detection efficiency $\eta$ and dark current ($DC$). The
data-taking time is expressed as $T$.
In the last years, preparation work has demonstrated the basics of the setup.

\section{Optics}

\begin{figure}
	\centerline{\includegraphics[width=0.8\textwidth]{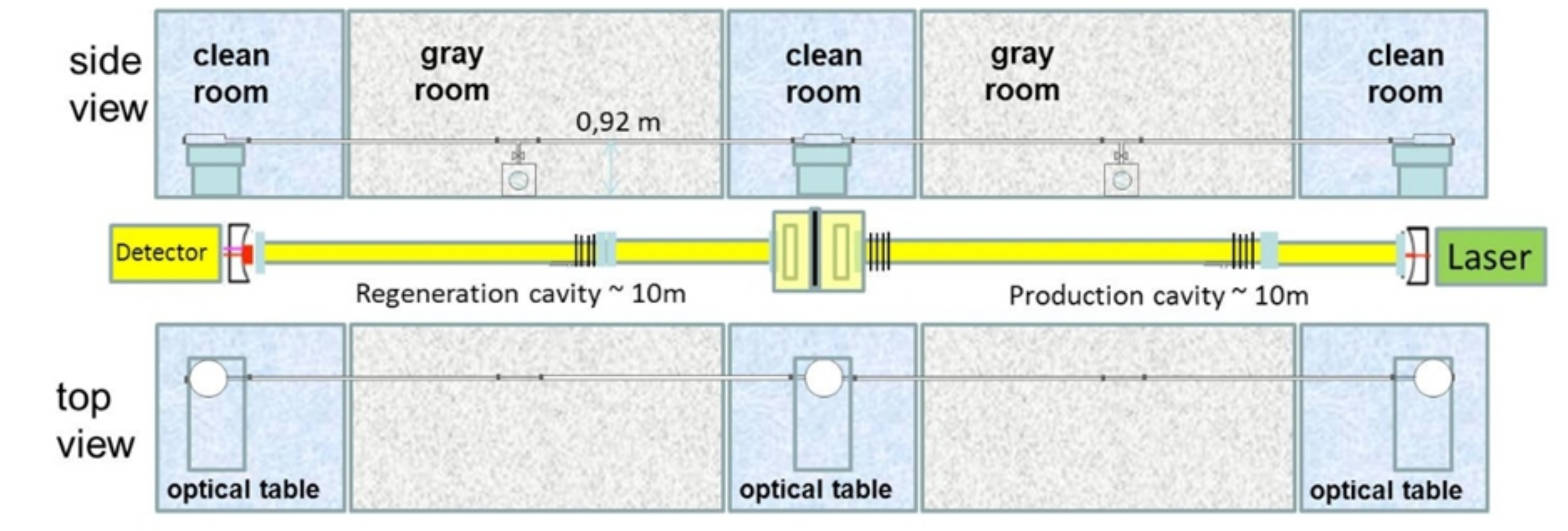}}
	\caption{ALPS IIa experiment.}\label{Fig:ALPSIIa}
	\label{sec:figures}
\end{figure}

The ALPS IIa (search for hidden photons) optical setup includes two 10 meters
optical cavities separated by a ligth-tight barrier. A 30~W 1064~nm laser is
injected inside the first cavity (Fig. \ref{Fig:ALPSIIa}). Such a system is technically challenging for two reasons: First, an alignment of both cavities towards each other
is necessary to provide a larger spatial overlap of the modes resonating in
both cavities. Second, high power buildups (PB) are required for both cavities
in order to reach the ALPS IIa foreseen sensitivity. The aimed PB of the
production cavity is of 5\,000 and the regeneration cavity PB is of
40\,000. In order to maximise this characteristic, the PC and RC need to be in
the same modal phase with a mode-overlap of 95~\%. The regeneration cavity is
locked via an auxiliary green beam obtained via second harmonic generation (KTP
crystal) of the PC infrared beam \cite{hodajerdi2015}.  Latest tests showed a
lower PB than required for the production cavity. Possible sources of such
issues are the mirrors coating, cleanliness of the mirrors, alignment of the
cavity as well as a clipping in the beam pipes. Usage of a cavity ring-down
technique demonstrated a good quality of the mirrors \cite{Isogai2013}.
Measurements will be repeated with a larger beam radius in order to enlarge the
tested region on the mirrors surface.

\section{Coupling of the beam inside a fiber}

The regeneration cavity will be connected via a fiber to a single-photon
detector in order to detect possible regenerated photons. Efficient coupling of a $4.23$~mm
beam inside a $8.2~\mu$m single-mode fiber is feasible but still needs to
be demonstrated to remain stable over longer time-scales.

\begin{figure}
	\includegraphics[width=7cm]{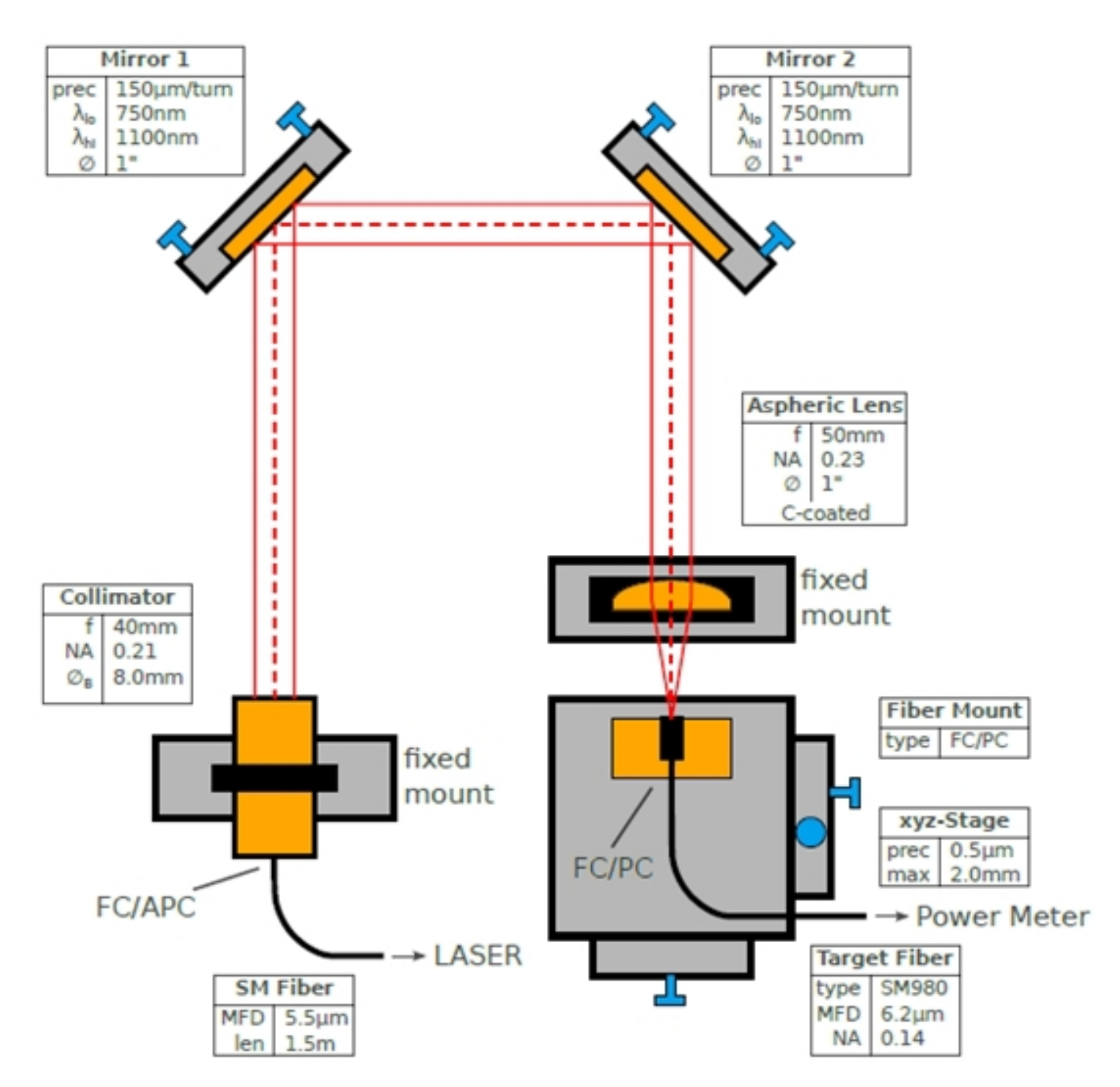}\hfill
	\includegraphics[width=7cm]{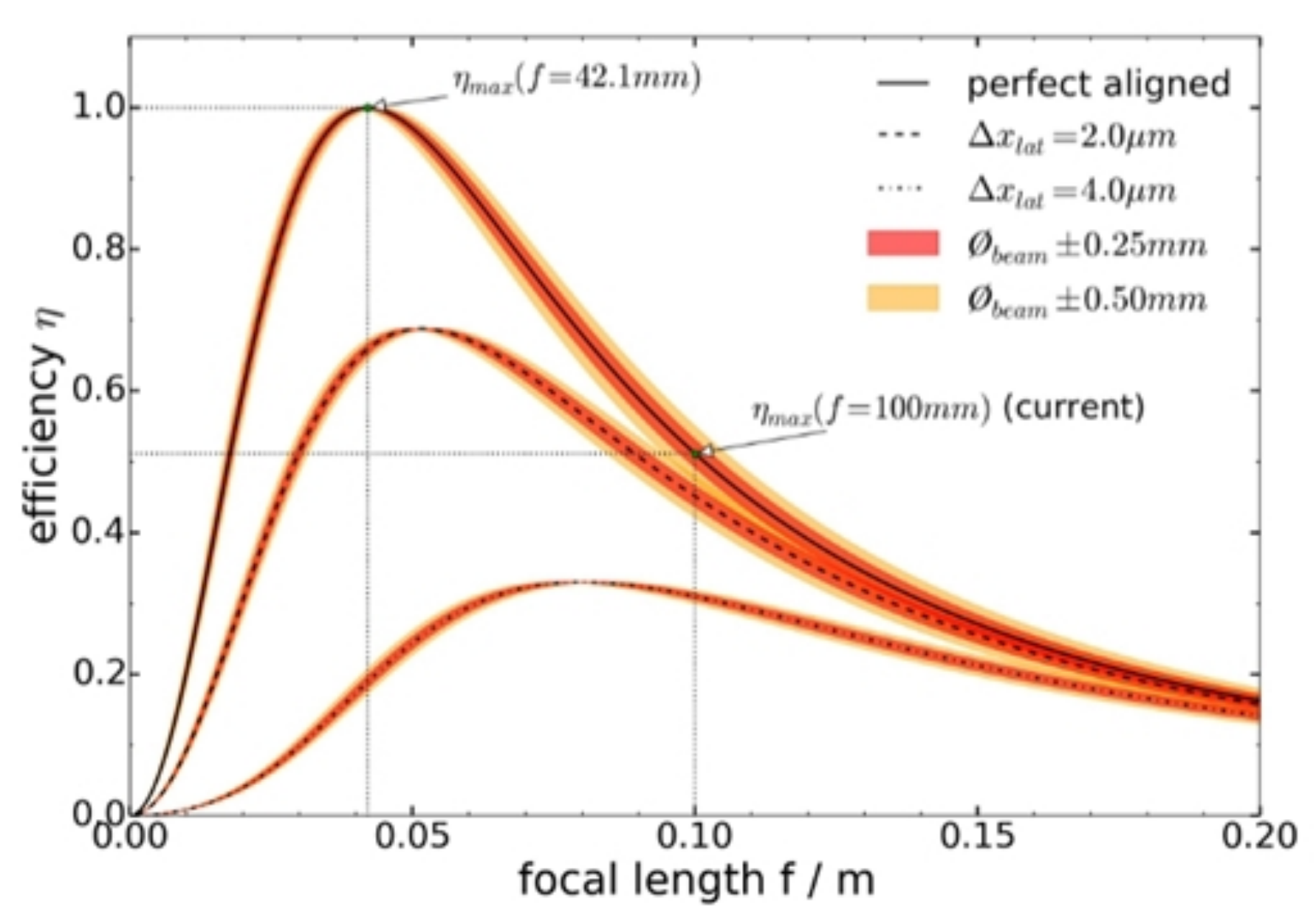}
	\caption{Coupling of the beam. On the left, a drawing of the coupling
of the beam test setup. On the right, the theoritical efficiency of the
coupling values $\eta$ for different levels of alignment $\Delta$$x_{tot}$ and
for different focal length f.}\label{Fig:Coupling} \end{figure}

The coupling of the beam inside a fiber setup includes two mirrors as well as
an aspheric lens (Fig. \ref{Fig:Coupling}).  In the test setup, a class 1 $\lambda=1064~$nm 
laser is shone to a mirror setup before being focused inside a standard
single-mode fiber.
It has been shown that the efficiency of the coupling depends highly on the
alignement of the setup and on the focal length of the used lens (Fig.
\ref{Fig:Coupling}). During the preliminary tests, an efficiency higher than 80
\% was reached. The highest value for the final setup which has been currently
obtained is of 53 \% for a focal length of 35~mm. This value is lower than what
was expected for such a lens. In near future, the beam quality will be studied
with a knife-edge unit. Such a device allows characterization and
adjustement of the beam on micrometer-scale before it enters the fiber.

\section{Detector}

The detection of a low rate (one event every few hours) of low energetic (1.17
eV) photons requires both, a high 
detection efficiency as well as a dark count rate. Additionally, the ALPS II
detection system is required to have a good energy and time resolution as well as a good long-term
stability. To meet all of these criteria, the ALPS II setup includes a 
cryogenic detector of the transition edge type (TES) developed
by NIST (National Institute of Standard and Technology) \cite{bastidonp2015}. 

Transition-Edge Sensors are superconductive microcalorimeters measuring the
temperature difference $\Delta$T induced by the absorption of a photon. The
positioning of the detector within its superconductive transition (30\% of its
normal resistance) is induced through a thermal link to a heat bath at
$T_{b}=80 mK$ and through a constant bias voltage. In order to obtain the
cool-down of the detector, it is placed in an adiabatic demagnetization
refrigerator (ADR) \cite{White2002}. 

The ALPS detector module includes two TESs inductively coupled to a SQUID
(Superconducting Quantum Interference Device). The ALPS detectors are optimized
for 1064~nm photons. The sensitive area of each chip measures 25 x 25
$\mu$$m^{2}$ for a thickness of 20 nm. The substrate is surrounded by a
standard fiber ceramic sleeve allowing connection of a single mode fiber
ferrule \cite{dreyling2015}.

NIST has demonstrated that such a detector can reach quantum efficiency higher
than 95~\% \cite{lita2008}. Latest measurements of the ALPS II detector
efficiency led to a first approximation of 30~\%. Optimization work is
currently under progress.

\section{ALPS IIc}

The ALPS IIc experiment will allow the search for axion-like particles (ALPs).
It is constituted in the same way as ALPS IIa with two 100 meters cavity and the
addition of 20 HERA (Hadron-Electron Ring Accelerator) dipoles \cite{Bahre2013}
to allow the conversion of photons into ALPs and re-conversion.  The HERA
dipoles were all bent during their design, leading to a small aperture of 35
mm. It was foreseen to unbend all of the dipoles by applying a force in their
middle (cold mass). 
The deformation of the first magnet was successful, yielding to an aperture of
50 mm allowing to set up the 100 m long cavities without any aperture
limitations. The magnet is working according to its specifications with a slight increase of its
quench current. Effort to straighten further magnets are on-going. 

\section{Summary}

The ALPS II experiment aims at an improvement of sensitivity by
a factor of 3\,000 compared to
ALPS I for the coupling of axion-like particles to photons. This improvement is achieved mainly by
implementing a regeneration cavity and a larger magnetic length. Basics of the
optics setup have been demonstrated but not all of the specifications have been
reached yet. A Tungsten Transition-Edge Sensor operated below 100 mK has been
successfully used to detect single-photons in the near-infrared.

\section{Acknowledgments}

The author would like to thank all the members of the ALPS collaboration. The author also thanks the PIER Helmholtz Graduate School for their financial travel support.
 
% ****************************************************************************
% BIBLIOGRAPHY AREA
% ****************************************************************************

\begin{footnotesize}

\end{footnotesize}

% ****************************************************************************
% END OF BIBLIOGRAPHY AREA
% ****************************************************************************

\end{document}